\documentclass[letterpaper,aps, pra,twocolumn, english, showpacs]{revtex4}

\usepackage{pslatex}
\usepackage[T1]{fontenc}
\usepackage[latin1]{inputenc}
\usepackage{amssymb, amsmath, amsfonts}

\providecommand{\LyX}{L\kern-.1667em\lower.25em\hbox{Y}\kern-.125emX\@}

\usepackage{babel}
\usepackage{graphicx}
\input{epsf}

\begin{document}
\title{Low-energy resonances and bound states of aligned bosonic and fermionic dipoles}
\author{K. Kanjilal and D. Blume}
\affiliation{Department of Physics and Astronomy, Washington State
University, Pullman, WA 99164-2814, USA}
\begin{abstract}
The low-energy scattering properties of two aligned identical bosonic and identical fermionic dipoles 
are
analyzed. Generalized scattering lengths are determined as functions of the dipole moment and the scattering energy. Near resonance, where a new bound state is being pulled in, all non-vanishing generalized scattering lengths diverge, with the $a_{00}$ and $a_{11}$ scattering lengths being dominant for identical bosons and identical fermions, respectively, near both broad and narrow resonances.  Implications for the energy spectrum and the eigenfunctions of trapped two-dipole systems and for 
pseudo-potential treatments
are discussed.  
\end{abstract}
\pacs{34.10.+x}
\maketitle

Currently, the creation of ultracold heteronuclear ground state molecules poses one of the major experimental challenges in the field of ultracold physics~\cite{coldmol}. The trapping of ultracold ground state molecules with large phase space density promises to allow an exciting array of novel research lines to be studied.
Although the largest phase space density of ultracold ground state  molecules achieved to date is still fairly small, a number of promising cooling schemes have been 
demonstrated~\cite{cooling}. 
Thus, it is expected that degenerate molecular gases with large electric dipole moment will be created in the laboratory in the near future. Polar molecules are a candidate for qubits in quantum computing~\cite{qc} and may be used in high precision measurements that aim at placing yet stricter limits on the electric dipole moment of the electron~\cite{edm}. Furthermore, dipolar gases are predicted to show roton-like features~\cite{roton} and to exhibit rich stability diagrams whose details depend on the trapping 
geometry~\cite{trapdep1}. 
The stability of dipolar 
atomic Cr condensates
has recently been investigated 
experimentally.
To enhance the anisotropic effects, 
which are due to Cr's magnetic dipole moment,
the $s$-wave scattering length was tuned to zero by applying an 
external field in the vicinity of 
a Fano-Feshbach 
resonance~\cite{beccr}.  

To create and then utilize ultracold molecules, it is mandatory to develop a detailed understanding of the scattering properties of two interacting dipoles in free space and in a trap. Unlike the interaction between $s$-wave alkali atoms, the interaction between two dipoles is long-range and angle-dependent. 
A two-dipole system can, e.g., be realized experimentally by loading an optical lattice with either two or zero dipoles per site. If the optical lattice is sufficiently deep and if the interaction between nearest and next to nearest neighbors are absent or negligible, then each optical lattice site can be treated as an independent approximately harmonic trap.

This paper determines the scattering properties of two aligned dipoles, either identical bosons or identical fermions, as functions of the dipole moment and the scattering energy. 
In general, the dipoles can either be magnetic or electric. For concreteness, 
we restrict our discussion in the following to the scattering between
molecular electric dipoles.
Sequences of scattering resonances, which can be classified as ``broad'' and ``narrow'', are found. For identical bosons, these resonances have previously been termed potential and shape resonances, respectively, and have been interpreted within the framework of adiabatic potential curves~\cite{bracket}. The resonance positions are correlated with the appearance of bound states in free space and ``diving'' states in the energy spectrum of two aligned dipoles under external confinement. The nature of the broad and narrow resonances is further elucidated by analyzing the bound state wavefunctions. In addition, we show that the eigenequation of two aligned dipoles in a harmonic trap interacting through an anisotropic zero-range pseudo-potential reproduces much of the positive energy spectrum, but exhibits some peculiar unphysical behavior for small energies. The origin of this unphysical behavior is pointed out and a simple procedure that eliminates it is presented.

Neglecting hyperfine interactions and treating each dipole as a point particle, the interaction potential between two dipoles aligned along the $z$-axis is for large interparticle distances $r$ given by $V_{\text{dd}}(\vec{r})$, $V_{\text{dd}}(\vec{r})=d^2(1-3\cos^2\theta)/r^3$. Here, $d$ denotes the dipole moment and $\theta$ the angle between the relative distance vector $\vec{r}$ and the $z$-axis. We model the short-range interaction $V_{\text{sr}}(\vec{r})$ between the dipoles by a simplistic hardwall potential, $V_{\text{sr}}(\vec{r})=\infty$ for $r < r_c$ and 0 for $r > r_c$, so that the full model potential is given by $V_{\text{m}}(\vec{r}) = V_{\text{sr}}(\vec{r})$ for $r < r_c$ and $V_{\text{dd}}(\vec{r})$ for $r > r_c$. The boundary condition imposed by $r_c$ can be thought of as introducing a short-range $K$-matrix, which is modified by the long-range dipole potential~\cite{debyou}. The characteristic length scale of $V_{\text{sr}}(\vec{r})$ is given by the hardcore radius $r_c$ and that of  $V_{\text{dd}}(\vec{r})$ by the dipole length $D_*$, $D_*=\mu d^2/\hbar^2$, where $\mu$ denotes the reduced mass. The corresponding natural energy scales are given by $E_{r_c}$ and $E_{D_*}$, respectively [$E_{r_c}=\hbar^2/(\mu r_c^2)$ and $E_{D_*}=\hbar^2/(\mu D_*^2)$]. A straightforward scaling of the relative Schr\"{o}dinger equation shows that $D_*$ and $r_c$ are not independent but that the properties of the system depend only on the ratio 
$D_*/r_c$~\cite{doertelong}. 
This ratio can be tuned experimentally by varying $D_*$ through the application of an electric 
field~\cite{debyou}.

To obtain the $K$-matrix elements $K_{l,m_{l}}^{l', m_{l'}}$, where $l$ and $l'$ denote the orbital angular momentum quantum number of the incoming and outgoing partial waves, respectively, and $m_l$ and $m_{l'}$ the corresponding projection quantum numbers, we solve the relative Schr\"{o}dinger equation for $V_{\text{m}}(\vec{r})$ for a fixed scattering energy $E_{\text{sc}}$ numerically. The azimuthal symmetry conserves the projection quantum number, and throughout we restrict our analysis to $m_l=0$. The radial Schr\"{o}dinger equation is propagated using the Johnson algorithm with adaptive step size~\cite{johnson}. The $K$-matrix elements $K_{l,0}^{l',0}(k)=\tan\delta_{l,l'}(k)$ are found by matching the log-derivative to the free-space solutions at sufficiently large $r$. Since the long-range part of $V_{\text{m}}(\vec{r})$ is proportional to the spherical harmonic $Y_{20}(\theta, \phi)$, the phase shifts $\delta_{l,l'}(k)$ are only non-zero if $|l-l'|\leq 2$.

\begin{figure}
\vspace{0.2cm}
\includegraphics[width=7cm]{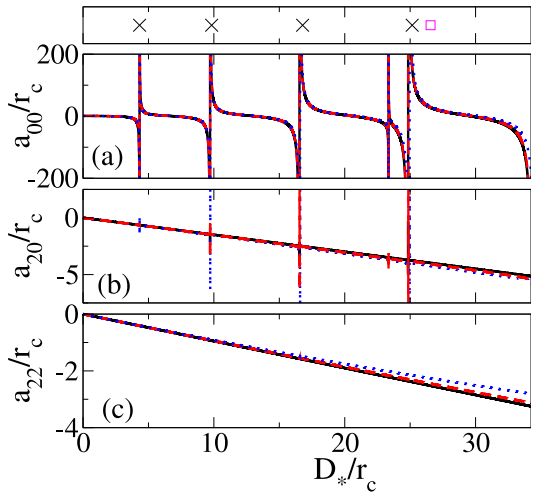}
\caption{\label{fig:one} Scaled scattering lengths (a) $a_{00}/r_c$, (b) $a_{20}/r_c$ and (c) $a_{22}/r_c$ as a function of the scaled dipole length $D_*/r_c$ for two identical bosons interacting through  
$V_{\text{m}}$ 
for three different scattering energies: $E_{\text{sc}}= 9.36  \times 10^{-8} E_{r_c}$ (solid line), $E_{\text{sc}}= 9.36 \times 10^{-6} E_{r_c}$ (dashed line) and $E_{\text{sc}}= 9.36 \times 10^{-5} E_{r_c}$ (dotted line). Crosses and squares in the top panel indicate the positions of the broad and narrow resonances, respectively, predicted by analyzing the WKB phase of the adiabatic potential curves (see text).} 
\end{figure}

\begin{figure}
\vspace{0.2cm}
\includegraphics[width=7cm]{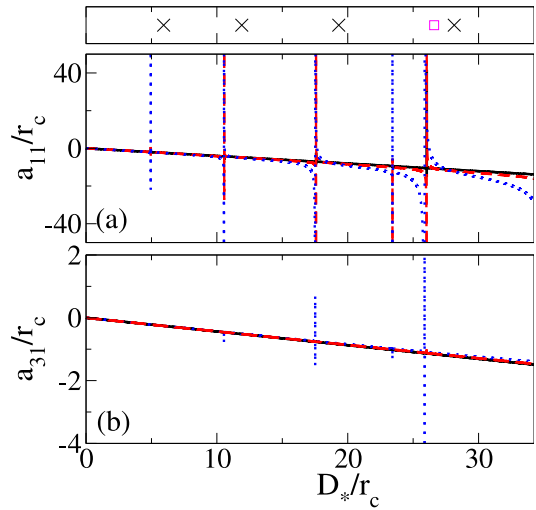}
\caption{\label{fig:two} Scaled scattering lengths (a) $a_{11}/r_c$ and (b) $a_{31}/r_c$ as a function of the scaled dipole length $D_*/r_c$ for two identical fermions interacting through 
$V_{\text{m}}$ 
for 
the
same three scattering energies as in Fig.~\ref{fig:one}. Crosses and squares in the top panel indicate the positions of the broad and narrow resonances, respectively, predicted by analyzing the WKB phase of the adiabatic potential curves (see text).} 
\end{figure}

Figures~\ref{fig:one} and~\ref{fig:two} show the generalized scattering lengths $a_{l,l'}$ for two identical bosons and two identical fermions, respectively, for three different scattering energies $E_{\text{sc}}$ as a function of the dipole length $D_*$. The scattering lengths $a_{l,l'}(k)$, $a_{l,l'}(k)=-K_{l,l'}(k)/k$ ($k$ denotes the wavevector, $k=\sqrt{2\mu E_{\text{sc}}/\hbar^2}$), are defined so that the $a_{l,l'}(k)$ approach a constant as $k\rightarrow 0$~\cite{threshold, yiyou}. The largest $D_*/r_c$ value considered in Fig.~\ref{fig:one} is 40. If we choose $r_c\approx 10 a_0 $, then the largest dipole length considered in Figs.~\ref{fig:one} and~\ref{fig:two} is $D_*^{\text{max}}\approx 400 a_0$, implying a minimum dipole energy $E_{D_*}^{\text{min}}$ of $1.27\times10^{-4}$ K. For the polar molecule OH, this corresponds to a maximum dipole moment of 0.404 Debye, a value that should be attainable experimentally. The scattering energies in Figs.~\ref{fig:one} and~\ref{fig:two} range from $9.36\times 10^{-8} E_{r_c}$ to $9.36\times 10^{-5} E_{r_c}$, or, using as before $r_c=10 a_0$, from $1.91\times 10^{-8}$ K to $1.91\times 10^{-5}$ K.  Thus, the largest $E_{\text{sc}}/E_{D_*}$ value considered in Figs.~\ref{fig:one} and~\ref{fig:two} is 0.15. This places the present study in the regime where the minimum value of the cross section has been predicted to behave 
universally~\cite{cavagnero},
but where the parameters of the two-body potential and the $s$-wave scattering 
length
it results in, especially near resonance, are important~\cite{ticknor}.

Figure~\ref{fig:one}(a) shows the scattering length $a_{00}$ as a function of $D_*$ for two identical bosons interacting through $V_{\text{m}}(\vec{r})$. Five broad and two narrow resonances (located at $D_*\approx 23 r_c$ and $37 r_c$) are clearly visible. Figures~\ref{fig:one}(b) and 
\ref{fig:one}(c) 
show the generalized scattering lengths $a_{20}$ and $a_{22}$, respectively. In the Born approximation 
(BA) 
for  $V_{\text{dd}}(\vec{r})$, both $a_{20}$ and $a_{22}$ vary linearly with $D_*$~\cite{yiyou, our}. $a_{20}$ and $a_{22}$ obtained from the full coupled channel calculation show deviations from the  
BA
for certain $D_*$ values. The positions of the ``spikes'' coincide with the resonance positions of $a_{00}$. Notably, the widths of the spikes decrease with increasing $l+l'$. The top panel of Fig.~\ref{fig:one} shows the resonance positions as predicted by the WKB phase accumulated in different adiabatic potential curves~\cite{bracket}. The crosses, obtained by analyzing the WKB phase of the lowest adiabatic potential curve, predict the positions of the broad resonances very accurately. The squares, obtained by summing the WKB phases of all other adiabatic potential curves, predict the number of narrow resonances semi-quantitatively but do not predict their positions  
accurately~\cite{bracket}.

Figures~\ref{fig:two}(a) and~\ref{fig:two}(b) show the generalized scattering lengths $a_{11}$ and $a_{31}$ for two aligned identical fermions interacting through $V_{\text{m}}(\vec{r})$ as a function of $D_*$. Away from resonance, $a_{11}$ and $a_{31}$ vary approximately linearly with $D_*$. The spikes in Fig.~\ref{fig:two} are interpreted as resonances, which we term, as in the boson case, broad and narrow \footnote{Strictly speaking, both types of resonances for identical fermions are narrow. However, since one type is narrower than the other, we refer to them as ``broad'' and ``narrow''.}. Figure~\ref{fig:two} shows five broad and two narrow resonances (located at $D_*\approx 23.5 r_c$ and $37.5 r_c$). A key difference between dipole scattering of identical bosons and identical fermions is that the lowest non-vanishing scattering length for bosons (i.e., $a_{00}$) cannot be approximated by applying the 
BA
to  $V_{\text{dd}}(\vec{r})$ (the 
BA
for  $V_{\text{dd}}(\vec{r})$ gives $a_{00}=0$) while the lowest non-vanishing scattering length for fermions (i.e., $a_{11}$) can be, away from resonance, approximated by the 
BA
for  $V_{\text{dd}}(\vec{r})$ (the 
BA
for  $V_{\text{dd}}(\vec{r})$ gives $a_{11}=-2D_*/5$)~\cite{yiyou, our}. The crosses and squares shown in the top panel of  Fig.~\ref{fig:two} indicate the positions of the broad and narrow resonances, respectively, as predicted from the WKB phase of the lowest adiabatic potential curve and of all other adiabatic potential curves. For identical fermions, the WKB prediction for the positions of the broad resonances is less accurate than that for identical bosons.

Figures~\ref{fig:one} and~\ref{fig:two} show that the widths of broad and narrow resonances increase with increasing $E_{\text{sc}}$ for fixed $D_*/r_c$ and with increasing $D_*/r_c$ for fixed $E_{\text{sc}}$. Putting this together, we find that the resonance widths of both broad and narrow resonances increase with increasing $E_{\text{sc}}/E_{D_*}$.

\begin{figure}
\vspace{0.2cm}
\includegraphics[width=7cm]{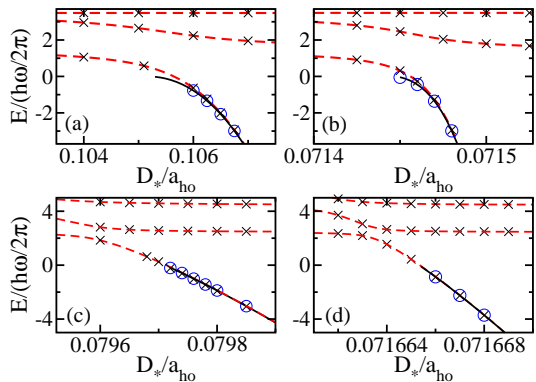}
\caption{\label{fig:eigen} Eigenenergies for two identical bosons in the vicinity of (a) a broad resonance and (b) a narrow resonance, and for two identical fermions in the vicinity of (c) a broad resonance and (d) a narrow resonance. 
Solid lines and circles show 
the
bound state energies for two dipoles in free space interacting through 
$V_{\text{m}}$ 
and  
$V_{\text{ps}}$, 
respectively. Dashed lines and crosses show the bound state energies for two dipoles under external harmonic confinement interacting through 
$V_{\text{m}}$ and $V_{\text{ps}}$, 
respectively. The oscillator length $a_{\text{ho}}$ is given by $\sqrt{\hbar/(\mu\omega})$, and the hard core radius $r_c$ of 
$V_{\text{m}}$ 
is 0.00306 $a_{\text{ho}}$.} 
\end{figure}

To better understand the resonance structure in Figs.~\ref{fig:one} and~\ref{fig:two}, we determine the bound state energies of the two interacting dipoles in free space. The Schr\"{o}dinger equation for the relative coordinate is solved using two-dimensional B-splines. The two-dipole system supports a new bound state at those $D_*/r_c$ values where the scattering lengths $a_{00}$ and $a_{11}$ for two identical bosons and fermions, respectively, diverge. Solid lines in Figs.~\ref{fig:eigen}(a) and~\ref{fig:eigen}(b) show the bound state energy for two identical bosons in the vicinity of a broad and a narrow resonance, respectively, while  solid lines in Figs.~\ref{fig:eigen}(c) and~\ref{fig:eigen}(d) show the bound state energy for two identical fermions in the vicinity of a broad and a narrow resonance, respectively. 

The two-body energy $E_b$ of weakly-bound $s$-wave interacting systems is well described by the $s$-wave scattering length $a_{00}$, $E_b=-\hbar^2/(2\mu [a_{00}(E)]^2)$. To test if this simple pseudo-potential expression holds for dipolar systems, we analytically continue the scattering lengths for $V_{\text{m}}(\vec{r})$ to negative energies. We obtain stable $a_{00}(E)$ and $a_{11}(E)$ for negative scattering energies by matching the coupled channel solutions to the free-space solutions 
at
relatively small $r$ values ($r_{\text{max}}\approx|k|^{-1}$). The bound state energies for two identical bosons in free space, determined 
self-consistently~\cite{a0e2,a0e} 
from 
$E_b=-\hbar^2/(2 \mu [a_{00}(E_b)]^2)$, 
are shown by circles in Figs.~\ref{fig:eigen}(a) and~\ref{fig:eigen}(b). Similarly, we determine the bound state energies for two identical fermions in free space by solving the equation $E_b=-\hbar^2/(2\mu [a_{11}(E_b)]^2)$~\cite{theorystock,bound} self-consistently [circles in Figs.~\ref{fig:eigen}(c) and~\ref{fig:eigen}(d)]. Somewhat surprisingly, the bound state energies in the vicinity of both broad and narrow resonances are very well described by a single-channel expression for identical bosons and identical 
fermions (see below for further discussion).

In addition to the free-space system, we consider the trapped system. Dashed lines in Fig.~\ref{fig:eigen} show the energies for two dipoles interacting through $V_{\text{m}}(\vec{r})$ under external harmonic confinement $V_{\text{trap}}$, $V_{\text{trap}}=\mu\omega^2r^2/2$. Near resonance, the lowest state with positive energy changes rapidly and turns into a negative energy state with molecular-like character. The energy of this ``diving'' state is slightly higher than the energy of the free-space system (the trap pushes the energy up).

\begin{figure}
\vspace{0.1cm}
\includegraphics[width=7cm]{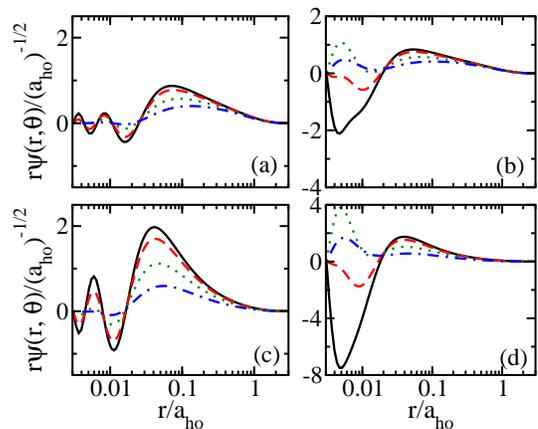}
\caption{\label{fig:wave} Eigenfunctions for two identical bosons in the vicinity of (a) a broad resonance ($D_*=0.1061 a_{\text{ho}}$) and (b) a narrow resonance ($D_*=0.07147 a_{\text{ho}}$), and for two identical fermions in the vicinity of (c) a broad resonance ($D_*=0.07976 a_{\text{ho}}$) and (d) a narrow resonance ($D_*=0.07167 a_{\text{ho}}$) in a spherical harmonic trap interacting via 
$V_{\text{m}}$ 
for  $\theta=0^{\circ}$ (solid), $\theta\approx18.2^{\circ}$ (dashed), $\theta\approx36.4^{\circ}$ (dotted) and $\theta\approx54.6^{\circ}$ (dash-dotted line). In all panels, $r_c$ equals 0.00306 $a_{\text{ho}}$. Note the log scale for $r$ and the different ranges of the $y$-axis.} 
\end{figure}

Figures~\ref{fig:wave}(a) and~\ref{fig:wave}(b) show the scaled eigenfunctions $r\psi(r, \theta)$ for two identical bosons with $E\approx-0.88 \hbar\omega$ as a function of $r$ for different $\theta$ 
near a broad and a narrow resonance, respectively. Similarly, Figs.~\ref{fig:wave}(c) and~\ref{fig:wave}(d) show the scaled eigenfunctions for two identical fermions with $E\approx-0.88 \hbar\omega$ 
near a broad and a narrow resonance, respectively. In all panels, the wavefunction cut for $\theta=0^{\circ}$ has the largest amplitude, reflecting the fact that the dipole-dipole potential is most attractive for $\theta=0^{\circ}$. Interestingly, the nodal structure of the wavefunction for two identical bosons near a broad resonance [Fig.~\ref{fig:wave}(a)] has a similar structure to that of the wavefunction of two identical fermions near a broad resonance [Fig.~\ref{fig:wave}(c)]: Both nodal surfaces show approximately spherical symmetry. On the other hand, the nodal structure of the wavefunction for two identical bosons near a narrow resonance [Fig.~\ref{fig:wave}(b)] has a similar structure to that of the wavefunction for two identical fermions near a narrow resonance [Fig.~\ref{fig:wave}(d)]: The nodal surfaces depend on both $r$ and $\theta$. 
To quantify the higher partial wave contributions,
we project the wave functions shown in Fig.~\ref{fig:wave}
onto spherical harmonics.
The $s$-wave contribution of the boson states 
near the 
broad and narrow resonances is about 95\%, while the $p$-wave contribution
of the fermion states near the broad and narrow resonances is about 95 and 80\%,
respectively. The gas-like states near resonance, in contrast, are dominated by a single partial wave (for bosons, e.g., the $s$-wave contribution of the energetically lowest-lying gas-like state is about 99\% while the $d$-wave contribution of the energetically next higher-lying state is about 99\%).
In the future, it will be interesting to investigate how the
higher partial wave contributions 
of the weakly-bound anisotropic molecules affect the scattering properties between two such composite particles and, more generally, the BEC-BCS crossover-type physics.

Lastly, we show that the entire energy spectrum of two aligned dipoles under external harmonic confinement interacting through $V_{\text{m}}(\vec{r})$ can be reproduced by a zero-range pseudo-potential framework. Since the anisotropy of the dipole-dipole interaction leads to a coupling 
of
different partial waves, the pseudo-potential $V_{\text{ps}}(\vec{r})$ contains an infinite number of terms~\cite{der}, $V_{\text{ps}}(\vec{r})=\sum_{l,l'=0}^{\infty}{g_{l,l'}(k)\Theta_{l,l'}(\vec{r})}$, where the coupling strength $g_{l,l'}(k)$ is proportional to $-\tan\delta_{l,l'}(k)/k^{l+l'+1}$ and $\Theta_{l,l'}(\vec{r})$ denotes an operator~\cite{der}. Assuming the $a_{l,l'}$ vanish for $|l-l'|>2$, as is the case for two interacting dipoles, the eigenequation for two particles under spherically symmetric external harmonic confinement can be elegantly written in terms of a continued fraction~\cite{our}. To obtain the eigenenergies for two aligned dipoles under external harmonic confinement interacting through $V_{\text{ps}}(\vec{r})$, we solve the eigenequation self-consistently, using the energy-dependent $a_{l,l'}(k)$ obtained for $V_{\text{m}}(\vec{r})$ as input parameters. The resulting energies, shown by crosses in Fig.~\ref{fig:eigen}, agree well with those obtained for $V_{\text{m}}(\vec{r})$ (dashed lines). However, for small $|E|$ the eigenequation for the pseudo-potential results in an unphysical eigenenergy [not shown in Figs.~\ref{fig:eigen}(a)-(d)]. For two identical bosons, e.g., the eigenequation for $V_{\text{ps}}(\vec{r})$ permits a solution with $E\approx 0.05\hbar\omega$, which is absent in the eigenspectrum of two identical bosons under external harmonic confinement interacting through $V_{\text{m}}(\vec{r})$. Importantly, if we restrict the pseudo-potential to the $V_{00}$ and $V_{11}$ terms for two identical bosons and fermions, respectively, the 
eigenspectra
for $V_{\text{ps}}(\vec{r})$ and
$V_{\text{m}}(\vec{r})$ agree very well for $E\lesssim0.5\hbar\omega$ 
(two identical bosons)
and $E\lesssim1.5\hbar\omega$  
(two identical fermions), 
and the unphysical eigenenergies are absent. This shows (i) that the scattering lengths $a_{00}$ (identical bosons) and $a_{11}$ (identical fermions) are dominant in this regime, and (ii) that the unphysical eigenenergies are due to the higher partial wave contributions of $V_{\text{ps}}(\vec{r})$. The latter can be understood as follows: The coupling strengths $g_{l,l'}(k)$ for two interacting dipoles are proportional to $a_{l,l'}(k)/k^{l+l'}$, and---since the $a_{l,l'}(k)$ are defined so that they approach a constant in the $k\rightarrow 0$ limit---diverge as $k$ goes to zero for $l+l'>0$. A detailed analysis of the eigenequation for $V_{\text{ps}}(\vec{r})$ shows that these divergences give rise to the unphysical eigenenergies for small $|k|$. No unphysical eigenenergies arise for larger $|k|$; in this regime, the $1/k^{l+l'}$ factor in $g_{l,l'}(k)$ can be thought of as a simple ``rescaling''. Furthermore, the unphysical eigenenergies do not arise if the phase shifts are obtained for a short-range model potential whose scattering lengths are defined by $-\tan\delta_{l,l'}(k)/k^{l+l'+1}$.
Although the pseudo-potential reproduces the eigenenergies well,
we note that the single-parameter description fails to describe the 
higher partial wave admixtures discussed in the context of Fig.~\ref{fig:wave}.

In summary,
this paper considers the scattering and bound state properties
of two interacting dipoles near 
resonance. Although our analysis has been performed 
for a simple model potential, we believe that the main conclusions 
hold more generally.
Near resonance, the magnitude of 
all non-vanishing scattering lengths becomes large, with $|a_{00}|$ being
largest for identical bosons and $|a_{11}|$ for two identical fermions.
We have found that the wave function of weakly-bound two-dipole systems 
contains higher partial wave contributions, raising interesting perspectives
for studying BEC-BCS crossover-type physics. 
Despite the admixture of higher partial waves, a single-parameter 
pseudo-potential treatment reproduces the
eigenenergy of the two-dipole system very accurately.

We acknowledge the support of the NSF through Grant No. PHY-0555316 and useful discussions with J. Bohn, Z. Idziaszek, G. Orso, Z. Pavlovic and H. Sadeghpour.

\end{document}